# The number of linked references of publications in Microsoft Academic in comparison with the Web of Science

Robin Haunschild*, Sven E. Hug**, Martin P. Brändle***, and Lutz Bornmann****

* Max Planck Institute for Solid State Research
Heisenbergstr. 1,
70569 Stuttgart, Germany.
Email: R.Haunschild@fkf.mpg.de

** Evaluation Office
University of Zurich
8006 Zurich, Switzerland

*** Zentrale Informatik and Main Library
University of Zurich
8006 Zurich, Switzerland

**** Division for Science and Innovation Studies
Administrative Headquarters of the Max Planck Society
Hofgartenstr. 8,
80539 Munich, Germany.
Email: bornmann@gv.mpg.de

Microsoft Academic (MA) is a promising new data source for bibliometrics (Harzing & Alakangas, 2017a; Hug & Brändle, 2017; Hug, Ochsner, & Brändle, 2017) but comprehensive studies on the quality of MA metadata are still lacking. However, there are six studies that evaluate various aspects of MA metadata beside their main research questions (Harzing, 2016; Harzing & Alakangas, 2017b; Herrmannova & Knoth, 2016; Hug & Brändle, 2017; Hug et al., 2017; Paszcza, 2016).

In the context of a comprehensive MA study, we investigated in a first exploration the quality of linked references data in MA. Linked references are the backbone of bibliometrics, because they are the basis of the times cited information in citation indexes. Furthermore, they are used to study networks of scientific publications (de Solla Price, 1965) and to normalize citations across sources (Bornmann & Haunschild, 2016; Glänzel & Thijs, 2017; Waltman & van Eck, 2013) and research fields (Waltman & van Eck, 2012). In the raw data of Scopus and the Web of Science (WoS), linked references are stored as database-internal IDs and cited references are available as bibliographic data, which allows for evaluating the accuracy of the provided bibliographic data and the performance of the employed citation matching algorithm (see Olensky, Schmidt, & van Eck, 2016).

MA, however, provides no other reference information than database-internal IDs of linked references (i.e. attribute 'RId' in the Academic Knowledge API[1]). There is one study that already touched on linked references in MA. Herrmannova and Knoth (2016) found that, in February 2016, 96.85 million of the total 126.91 million publications in MA had zero linked references (i.e. 76.3% of all records in MA). This high number of zero linked references questions the quality of cited references data in MA. In this Letter to the Editor, we explore further aspects of linked references in MA to assess MA's quality as a data source for bibliometrics. In particular, we compare the number of linked references in MA and WoS on the level of individual publications.

---

[1] https://www.aka.ms/AcademicAPI

Our publication set was drawn from the Zurich Open Repository and Archive (ZORA), an open archive and repository in which the University of Zurich documents its complete publication output since 2008. The coverage of the ZORA records from 2008 to 2016 (83,663 items) was checked in MA and WoS as described by Hug and Brändle (2017). In total, 34,774 ZORA records are covered by both databases. Metadata from MA were retrieved via the Academic Knowledge API (AK API) as described by Hug and Brändle (2017). The WoS data used for this Letter to the Editor are from an in-house database developed and maintained by the Max Planck Digital Library (MPDL, Munich) and derived from the Science Citation Index Expanded (SCI-E), Social Sciences Citation Index (SSCI), Arts and Humanities Citation Index (AHCI) provided by Clarivate Analytics (Philadelphia, Pennsylvania, USA). The in-house database covers a snapshot of the WoS database from the last week of April 2017. MA data was collected in the first week of June 2017.

Out of the 34,774 publications covered by both databases, 2,836 have zero linked references in MA (8.2%) but no publication has zero linked references in WoS. Surprisingly, none of the publications from MA has more than 50 linked references. Inquiring the development team of MA on this issue, they explained that the number of linked references per publication that can be retrieved via the AK API is limited to 50 in order to keep the API performant (A. Chen, personal communication, June 9, 2017). Hence, we restricted the comparison of MA and WoS to those publications which have less than 50 linked references in MA (i.e., 25,539 items). Out of these 25,539 papers, 2,836 (11.1%) have zero linked references although the number of linked references is greater than zero in the WoS for each of those papers (on average, 34.5 linked references per paper). The concordance coefficient according to Lin (1989) between the number of linked references of all analyzed MA and WoS records is 0.10. Following the interpretation rules of Koch and Sporl (2007), the concordance between linked references in MA and WoS is between weak and non-existent. In fact, this low concordance between MA and WoS data originates from papers in our sample (25,539 publications) which have 50 or more linked references in WoS (751 publications). Restricting the publication set to papers with less than 50 linked references in WoS (n=24,788) results in a concordance coefficient of 0.68 which indicates a strong agreement.

Linked references are an important data source for several bibliometric analyses (as outlined above). Since we reveal in our exploration that the MA data are limited to maximal 50 linked references per paper, we asked Microsoft whether the threshold of 50 is used for the calculation of the times cited information. They answered that the limit is not applied when the number of citations (attribute CC in the AK API) is being calculated in MA (D. Eide, personal communication, June 12, 2017). MA citation counts are based on the 'unlimited' number of linked references.

Based on our first exploration of MA data, we suggest that Microsoft publishes the limits they impose on the data in the AK API. Furthermore, we recommend that Microsoft expands the number of references available in the API as much as possible. Actually, any limit on the number of linked references might pose a potential problem for usage of MA data in research evaluation. Therefore, no limit at all would be preferable.

The comparison of the number of cited references from MA and WoS was the first exploration in our project. We will go on with the project by comparing MA data with the data from other databases and exploring them for field- and time-normalization as well as bibliometric networking. We hope other bibliometricians will follow by analyzing this new data source.

# References

Bornmann, L., & Haunschild, R. (2016). Citation score normalized by cited references (CSNCR): The introduction of a new citation impact indicator. *Journal of Informetrics, 10*(3), 875-887. doi:10.1016/j.joi.2016.07.002

de Solla Price, D. J. (1965). Networks of Scientific Papers. *Science, 149*(3683), 510-515. doi:10.1126/science.149.3683.510

Glänzel, W., & Thijs, B. (2017). *Bridging another gap between research assessment and information retrieval – The delineation of document environments*. Paper presented at the STI 2017, Paris.

Harzing, A.-W. (2016). Microsoft Academic (Search): a Phoenix arisen from the ashes? *Scientometrics, 108*(3), 1637-1647. doi:10.1007/s11192-016-2026-y

Harzing, A.-W., & Alakangas, S. (2017a). Microsoft Academic is one year old: the Phoenix is ready to leave the nest. *Scientometrics, 112*(3), 1887-1894. doi:10.1007/s11192-017-2454-3

Harzing, A.-W., & Alakangas, S. (2017b). Microsoft Academic: is the phoenix getting wings? *Scientometrics, 110*(1), 371-383. doi:10.1007/s11192-016-2185-x

Herrmannova, D., & Knoth, P. (2016). An analysis of the Microsoft Academic Graph. *D-Lib Magazine, 22*(9/10). doi:10.1045/september2016-herrmannova

Hug, S. E., & Brändle, M. P. (2017). The coverage of Microsoft Academic: Analyzing the publication output of a university. *Scientometrics*. doi:10.1007/s11192-017-2535-3

Hug, S. E., Ochsner, M., & Brändle, M. P. (2017). Citation analysis with Microsoft Academic. *Scientometrics*. doi:10.1007/s11192-017-2247-8

Koch, R., & Sporl, E. (2007). Statistical methods for comparison of two measuring procedures and for calibration: Analysis of concordance, correlation and regression in the case of measuring intraocular pressure. *Klinische Monatsblatter Fur Augenheilkunde, 224*(1), 52-57. doi:10.1055/s-2006-927278

Lin, L. I. (1989). A CONCORDANCE CORRELATION-COEFFICIENT TO EVALUATE REPRODUCIBILITY. *Biometrics, 45*(1), 255-268. doi:10.2307/2532051

Olensky, M., Schmidt, M., & van Eck, N. J. (2016). Evaluation of the Citation Matching Algorithms of CWTS and iFQ in Comparison to the Web of Science. *Journal of the Association for Information Science and Technology, 67*(10), 2550-2564. doi:10.1002/asi.23590

Paszcza, B. (2016). *Comparison of Microsoft Academic (Graph) with Web of Science, Scopus and Google Scholar.* (Master's Thesis), University of Southampton, Southampton. Retrieved from http://eprints.soton.ac.uk/id/eprint/408647

Waltman, L., & van Eck, N. J. (2012). A new methodology for constructing a publication-level classification system of science. *Journal of the American Society for Information Science and Technology, 63*(12), 2378-2392. doi:10.1002/asi.22748

Waltman, L., & van Eck, N. J. (2013). Source normalized indicators of citation impact: an overview of different approaches and an empirical comparison. *Scientometrics, 96*(3), 699-716. doi:10.1007/s11192-012-0913-4